\documentclass[authoryear,12pt,letter,3p]{jowarticle}
\usepackage{graphicx}
\usepackage{amsmath}
\usepackage{amssymb}
\usepackage{natbib}
\usepackage{setspace}
\usepackage[all]{xy}
\makeatletter
\renewcommand\section{\@startsection{section}{1}{\z@}{-3.25ex plus -1ex minus -.2ex}{1.5ex plus .2ex}{\normalsize\bf}}
\renewcommand\subsection{\@startsection{subsection}{2}{\z@}{-3.25ex plus -1ex minus -.2ex}{1.5ex plus .2ex}{\normalsize\bf}}
\renewcommand\subsubsection{\@startsection{subsubsection}{3}{\z@}{-3.25ex plus -1ex minus -.2ex}{1.5ex plus .2ex}{\normalsize\bf}}
\makeatother

\newtheorem{thm}{Theorem}

\newcommand{\supp}[1]{\text{supp}(#1)}

\begin{document}
\begin{frontmatter}
\title{Conservation, Inertia, and Spacetime Geometry}

\author{James Owen Weatherall}\ead{weatherj@uci.edu}
\address{Department of Logic and Philosophy of Science\\ University of California, Irvine}
\begin{abstract}As Harvey Brown emphasizes in his book \emph{Physical Relativity}, inertial motion in general relativity is best understood as a theorem, and not a postulate.  Here I discuss the status of the ``conservation condition'', which states that the energy-momentum tensor associated with non-interacting matter is covariantly divergence-free, in connection with such theorems.  I argue that the conservation condition is best understood as a consequence of the differential equations governing the evolution of matter in general relativity and many other theories.  I conclude by discussing what it means to posit a certain spacetime geometry and the relationship between that geometry and the dynamical properties of matter.\end{abstract}
\end{frontmatter}
\doublespacing
\section{Introduction}\label{introduction}

In \emph{Physical Relativity} \citep{Brown}, Harvey Brown defends a number of startling and provocative claims regarding the status of inertial motion in general relativity and other theories.\footnote{See also \citet{Brown+Pooley}.}  He writes, for instance, that,
 \begin{quote}\singlespacing Inertia, before Einstein's general theory of relativity, was a miracle.  I do not mean the existence of inertial mass, but the postulate that force-free (henceforth \emph{free}) bodies conspire to move in straight lines at uniform speeds while being unable, by \emph{fiat}, to communicate with each other.  It is probably fair to say that anyone who is not amazed by this conspiracy has not understood it.  (And the coin has two sides: anyone who is not struck by the manner in which the general theory is able to explain the conspiracy ... has failed to appreciate its strength....) (pp. 14-5)\end{quote}
 Later, he clarifies the sense in which general relativity (GR) is distinctive.  He writes, ``GR is the first in the long line of dynamical theories, based on that profound Aristotelian distinction between natural and forced motions of bodies, that \emph{explains} inertial motion'' (p. 141).  GR achieves this explanatory virtue by dint of the fact that whereas in ``...Newtonian mechanics and SR [special relativity], the conspiracy of inertia is a postulate...,'' in GR, ``...the geodesic principle [i.e., the principle governing inertial motion] is no longer a postulate but a theorem'' (p. 141).

It is true that there are theorems in general relativity that capture the geodesic principle.  (In what follows, I will focus on one such result, due to \citet{Geroch+Jang}.)\footnote{\label{GPT} In addition to the Geroch-Jang theorem, there are also the results of \citet{Souriau}, \citet{Sternberg+Guillemin}, \citet{Ehlers+Geroch}, \citet{Wald+Gralla}, \citet{Geroch+Weatherall}, and others.  The so-called ``problem of motion'' in general relativity has a long history, going back to \citet{Einstein+Grommer} and \citet{Einstein+etal}.  \citet{Havas}, \citet{Kennefick}, and \citet{LehmkuhlHybrid} discuss aspects of this history, while \citet{LehmkuhlCareful} describes some of the key differences between Einstein's approach to the problem and modern approaches.  For surveys of the now-vast literature on the subject, see  \citet{Damour}, \citet{Blanchet}, \citet{Poisson}, \citet{Asada}, and the contributions to \citet{Puetzfeld+etal}; see also \citet{Tamir}, who discusses several approaches but argues against the claim that any of these really establish geodesic motion for actual small bodies.  \citet{Geroch+Weatherall} clarify the relationship between Ehlers-Geroch-Jang-style ``curve first'' results and results based on distributions, such as are developed by \citet{Sternberg+Guillemin}, strengthening both classes of result.  Although in my view, a number of recent results improve on the Geroch-Jang theorem, the Geroch-Jang theorem is nonetheless attractive for foundational discussions because it is simple, perspicuous, and by now familiar to many philosophers of physics because it has been the focus of several papers discussing inertial motion in connection with Brown's work.}   Even so, Brown's arguments concerning the special status of inertial motion in GR have been controversial.  For instance, Brown places special weight on the fact that it is a consequence of Einstein's equation that energy-momentum must be (locally) conserved, in the sense that the covariant divergence of the total energy-momentum of (source) matter must vanish.  This conservation condition plays an essential role in the theorems governing inertial motion in general relativity.  And so, Brown argues, inertial motion in GR follows from Einstein's equation.  But as \citet{MalamentGP} emphasizes, conservation of energy-momentum alone is not sufficient to establish the geodesic principle.  He shows that one also requires an energy condition, effectively requiring that energy-momentum is positive in a suitable sense.  Malament goes on to show that in the absence of any energy condition, conserved matter may follow any timelike curve at all.\footnote{Regarding just what energy condition is needed to prove the Geroch-Jang theorem, see \citet{WeatherallEC}; \citet{Bezares+etal} generalize the Ehlers-Geroch strengthening of the Geroch-Jang theorem, using a weaker averaged energy condition.  For more on energy conditions in general, included extensive discussion of their status and interpretation, see \citet{CurielEC}.}

Another line of argument puts pressure on the claim that GR is sharply distinguished from other theories with regard to inertial motion, with a special focus on Newtonian gravitation.  In particular, one can prove a theorem in the context of the geometrized reformulation of Newtonian gravitation (GNG), sometimes known as Newton-Cartan theory, that is strikingly similar to the Geroch-Jang theorem \citep{WeatherallMPNT}.\footnote{\label{GNGrefs} Note that although the theorem is set in GNG, it also covers ordinary Newtonian gravitation, which corresponds in this context to the special case where spacetime happens to be flat.  (In the same sense, the Geroch-Jang theorem may be understood to cover special relativity, a point I will return to below.)  GNG is of interest for several reasons.  One reason is that it casts Newtonian gravitation in a language maximally similar to GR, facilitating comparative analyses. Another reason is that there is a natural sense in which it is the ``classical limit'' of GR, as discussed by \citet{EhlersLimit1, EhlersLimit2}, \citet{Kunzle}, \citet{MalamentLimit1,MalamentLimit2}, \citet{WeatherallExplanation}, and \citet{FletcherLimit}.  For more on GNG and its relationship to Newtonian gravitation, see \citet{Trautman}, \citet{GlymourTE}, \citet{Friedman}, \citet{Bain}, \citet{MalamentGR}, \citet{Knox}, and \citet{WeatherallTE}.  The status of inertial motion in Newtonian gravitation is also discussed in detail by \citet{Earman+Friedman}, who anticipate the claim that it may be understood as a theorem.}  I have argued elsewhere that, in light of this theorem, the status of inertial motion in Newtonian gravitation is strikingly similar to the situation in GR \citep{WeatherallSGP, WeatherallLehmkuhl}.  In both theories, if one assumes an energy condition (of, arguably, similar strength in each case) and a conservation condition (of the same form), the geodesic principle becomes a theorem.  I have also argued that there is thus a sense in which \emph{both} theories may be said to explain inertial motion, equally well---though the conditions required in the theorems mean that the sense of explanation involved is subtle \citep{WeatherallLehmkuhl}.\footnote{This sense of explanation is related to what I have previously called the ``puzzleball view'' of theories.  \citet{Samaroo} has recently argued against taking either of these theorems to provide \emph{explanations}; in both cases, he takes them to provide \emph{explications} of inertial motion.  I do not want to quibble over these words, and I tend to agree with the thrust of his arguments.  \citet{Sus}, meanwhile, has maintained that despite the result discussed above, it is nonetheless the case that the explanation one gets in GR is ``more dynamical''.  I will return to these arguments in section \ref{conclusion}.}

Although the role of energy conditions in these theorems is interesting and important, I will set that issue aside in what follows.  Instead, my focus will be on what Brown himself suggests is the heart of the matter: the status of the conservation condition stating that the energy-momentum (or, in GNG, mass-momentum) of a free body is divergence-free. As Brown emphasizes, in GR this conservation condition must hold for the total source matter appearing in Einstein's equation.  Nothing analogous can be claimed in GNG, where the geometrized Poisson equation---the analog to Einstein's equation---does not imply anything about the divergence of the mass-momentum tensor.  And so, in GNG the conservation condition must be assumed independently.  I have previously argued both that Brown's claim about Einstein's equation, though true, is tangential to the point at hand, and also that the conservation condition should be understood as a meta-principle, expected to hold in a wide range of spacetime theories.

I still endorse both of these arguments.  But I think there is more to say.  In what follows, I will echo and amplify an observation due to \citet[pp. 543]{Pooley}, to the effect that the conservation condition is doubly secured in GR: it follows both from Einstein's equation and also from the differential equations governing matter.\footnote{Pooley cites a classic paper by \citet{TrautmanCLGR} to support this claim; Trautman's justification is ultimately very close to the view defended here.}  That is, it is a general property of the class of differential equations that describe matter in GR that, associated with any solution to those equations, there is always a quantity with the relevant properties of an energy-momentum tensor that is divergence-free whenever the matter in question is non-interacting. This fact follows from the role that the spacetime metric and derivative operator play in those equations.  More importantly, it holds much more generally than just in GR, including in theories where the analogue of Einstein's equation does not imply that the energy-momentum tensor is divergence-free (including GNG), or where there is no analogue to Einstein's equation because spacetime structure is fixed (as in SR or Newtonian gravitation). This situation, I will argue, indicates a deep link between conservation and inertia, on the one hand, and the ambient geometry in which a system of differential equations is set.\footnote{One can take these arguments to generally expand on my earlier claim that conservation is a ``meta-principle''.  The idea here is to show why we should expect it to hold so generally.}

The remainder of the paper will proceed as follows.  I will begin by stating and briefly discussing the Geroch-Jang theorem.  As I will argue, the important question is not, ``is energy-momentum divergence-free?'' but rather, ``with respect to what derivative operator is energy-momentum divergence-free?''  Next I will discuss the status of the conservation condition in GR, showing the sense in which it follows from the class of differential equations governing matter.  The technical arguments here are well known, and in fact Brown discusses them briefly in an appendix to \emph{Physical Relativity}.  But there is some value to rehearsing them, and to drawing the connection between what does the work in these arguments and what is necessary for the Geroch-Jang theorem.  Next I will turn to some alternative relativistic theories in which the analogue to Einstein's equation does not secure conservation, including Unimodular gravity, Tensor-Vector-Scalar gravity (TeVeS), and SR, showing how conservation---and hence, inertial motion---nonetheless arises in each of these cases. I will next turn to Newtonian gravitation, presenting the analogue of the Geroch-Jang theorem in that context and showing that there, too, one can expect that there will always be a conserved quantity associated with matter.  I will conclude by drawing morals concerning the relationship between the dynamics of matter and spacetime geometry.

\section{The Geroch-Jang Theorem}

In what follows, a \emph{relativistic spacetime} is a pair $(M,g_{ab})$, where $M$ is a smooth, four-dimensional, connected manifold, assumed to be Hausdorff and second countable, and $g_{ab}$ is a smooth, Lorentzian metric on $M$.\footnote{\label{GRbooks} I work in the signature $(1,-1,-1,-1)$.  In what follows, I will generally assume that all of the objects---manifolds, tensor fields, etc.---that are candidates to be smooth are smooth.  I will also limit attention to tangent structures, so that vectors, vector fields, and so forth should all be understood as tangent to the manifold.  For background on either general relativity or Lorentzian geometry, the reader is encouraged to consult \citet{Hawking+Ellis}, \citet{Wald}, \citet{ONeill}, and \citet{MalamentGR}.}  Associated with any relativistic spacetime is a unique torsion-free derivative operator $\nabla$ satisfying the condition $\nabla_a g_{bc}=\mathbf{0}$, known as the Levi-Civita derivative operator.  Any derivative operator determines a class of privileged curves, known as \emph{geodesics}.  A geodesic of a derivative operator is a curve $\gamma$ whose tangent field $\xi^a$ satisfies $\xi^n\nabla_n\xi^a=\mathbf{0}$; such curves are naturally understood as non-accelerating, by the standard of acceleration set by $\nabla$. In what follows, we will say that a (tangent) vector $\xi^a$ at a point is \emph{timelike} if $g_{ab}\xi^a\xi^b > 0$.

The theorem may then be stated as follows.\footnote{This statement of the theorem follows \citet[Prop. 2.5.2]{MalamentGR} closely.  See fn. \ref{GPT} for references to other results and discussions.}  (Here and in the rest of the next two sections, all indices are raised and lowered using $g_{ab}$.)
\begin{thm}\label{GJ}\singlespacing
\emph{(\textbf{\citet{Geroch+Jang}})}
Let $(M,g_{ab})$ be a relativistic spacetime.  Let $\gamma:I\rightarrow M$ be a smooth, imbedded curve.  Suppose that given any open subset $O$ of $M$ containing $\gamma[I]$, there exists a smooth, symmetric field $T^{ab}$ with the following properties.
\begin{enumerate}
\item \label{sdec} $T^{ab}$ satisfies the \emph{strengthened dominant energy condition}, i.e. given any timelike vector $\xi^a$ at any point in $M$, $T^{ab}\xi_a\xi_b\geq 0 $ and either $T^{ab}=\mathbf{0}$ or $T^{ab}\xi_b$ is timelike;
\item \label{cons}$T^{ab}$ satisfies the \emph{conservation condition}, i.e. $\nabla_b T^{ab}=\mathbf{0}$;
\item \label{inside}$\supp{T^{ab}}\subset O$; and
\item \label{non-vanishing}there is at least one point in $O$ at which $T^{ab}\neq \mathbf{0}$.
\end{enumerate}
Then $\gamma$ is a timelike curve that may be reparametrized as a geodesic.
\end{thm}

Several remarks are in order concerning the interpretation of this theorem.  First, in GR and in all of the other physical theories discussed in this paper, the points of $M$ represent events in space and time.  Matter is represented by fields on $M$, i.e., smooth assignments of mathematical objects to points of $M$.  There are different kinds of matter that one might encounter---perfect fluids, electromagnetic fields, Klein-Gordon fields, and so on---but matter of any kind is associated with a smooth, symmetric, rank-2 tensor, $T^{ab}$, representing that matter's energy and momentum densities as determined by arbitrary observers.\footnote{For discussions of how $T^{ab}$ encodes this information, see \citet[\S3.2]{Hawking+Ellis}, \citet[\S3.1]{Sachs+Wu}, and \citet[\S2.5]{MalamentGR}.}  This tensor is generally presumed to vanish in an open region if and only if there is no matter present in that region, reflecting the assumptions that (1) all matter is associated with some quantity of energy and momentum, and (2) all energy and momentum is associated with matter.\footnote{Cf. \citet[p. 61]{Hawking+Ellis}.  Of course, there is a controversial issue related to whether vacuum phenomena, such as gravitational waves, should be associated with energy in some extended sense.  I set that aside here.  At very least, all energy and momentum of the sort that may be encoded in an energy-momentum tensor is associated with matter.}  Thus, the $T^{ab}$ fields appearing in the Geroch-Jang theorem may be taken to characterize configurations of matter (of a generic, non-specified sort) that is localized to regions near a curve $\gamma$.

This matter is presumed to satisfy two conditions.  The strengthened dominant energy condition captures a sense in which the mass-energy associated with the matter is always positive and also that the 4-momentum density determined by any observer is a timelike vector, which may be understood as a causality condition satisfied by massive matter.\footnote{\label{EC} Note that pure radiation will generally not satisfy the strengthened dominant energy condition.  As noted in the introduction, some energy condition or other is needed in order to get geodesic motion, and for the theorem as stated, where the fact that $\gamma$ is timelike is a consequence of the theorem, various well-known weaker conditions will not suffice \citep{MalamentGP,WeatherallEC}.  On the other hand, if one begins by supposing that $\gamma$ is a timelike curve, it seems there is much more flexibility concerning the energy condition \citep{Geroch+Weatherall}.}  And the conservation condition captures a sense in which the matter in question is \emph{free}, i.e., not interacting with any other matter.  (Of course, much more will be said about the conservation condition and its interpretation in what follows.)  Both of these conditions are supposed to hold of generic, physically reasonable, massive matter that is not interacting with its environment.

Thus, the theorem may be interpreted as follows.  If a curve $\gamma$ is such that for any neighborhood of that curve, there exists some configuration of generic, physically reasonable, free, massive matter, which is present (only) in that neighborhood, then the curve must be a timelike geodesic.  In other words, the only curves along which free, arbitrarily small ``bodies,'' i.e., spatially compact matter configurations, may propagate are timelike geodesics.  Under this interpretation, the theorem may be seen to capture the \emph{geodesic principle}, which states that in the absence of external forces, massive test point particles follow timelike geodesics.\footnote{Note that there is a small gap, here.  The Geroch-Jang theorem says that the \emph{only} curves along which arbitrarily small, free, massive bodies may propagate are timelike geodesics; it does not say that there are any bodies that do so propagate, in the sense that it does not establish that any curves satisfy the antecedent of the theorem.  In fact, however, one can construct perfect fluids satisfying the conditions of the theorem in arbitrarily small neighborhoods of timelike geodesics in arbitrary spacetimes.}  Here we understand a ``massive particle'' as the small-body limit of generic, physically reasonable massive matter, in the sense described above.  Note that the geodesic principle, as stated here, concerns \emph{test} matter (as opposed to [gravitational] \emph{source} matter), in the sense that we do not suppose that the metric $g_{ab}$ depends on the presence of the particle.  This is reflected in the Geroch-Jang theorem by the fact that it presumes a fixed background spacetime $(M,g_{ab})$, and one does not place any constraints on the relationship between that spacetime and any of the $T^{ab}$ fields under consideration.\footnote{\citet{Ehlers+Geroch} strengthen the Geroch-Jang theorem to show that even if one considers extended bodies that \emph{do} contribute to $g_{ab}$ via Einstein's equation, then in the small-body limit these bodies follow timelike geodesics as long as their contribution to a background spacetime metric becomes arbitrarily small.}

This last point has an important consequence.  As will be discussed in the next section, any relativistic spacetime may be understood as a possible universe allowed by GR, where Einstein's equation asserts that the total energy-momentum tensor $T^{ab}$ arising from all matter in the universe is proportional to the \emph{Einstein tensor}, $G^{ab}$, defined by:
\begin{equation}\label{EinsteinTensor}
G_{ab} = R_{ab} - \frac{1}{2}Rg_{ab},\end{equation} where $R_{ab}=R^n{}_{abn}$ is the Ricci curvature associated with $\nabla$ and $R=R^a{}_a$ is the curvature scalar.   But as I just noted, neither Einstein's equation nor the Einstein tensor play any role in the Geroch-Jang theorem.  And so the theorem does not require one to understand the spacetime $(M,g_{ab})$ within the context of any particular physical theory.  Instead, it captures a general fact about curves on Lorentzian manifolds.  This means that it is readily applied to any physical theory that takes spacetime to be represented by a Lorentzian manifold, irrespective of any relationship that might obtain between the metric on that manifold and the energy-momentum density of matter, so long as matter in that theory satisfies the conservation and energy conditions.  Indeed, insofar as these conditions really do reflect generic facts about \emph{matter}, the Geroch-Jang theorem reveals not only that the geodesic principle is considerably more general than GR itself, but that the status of the geodesic principle as a \emph{theorem}, not a postulate, is similarly general.

Regarding this generality, there is another feature of the Geroch-Jang theorem that is worth emphasizing.  The theorem establishes a relationship between two classes of objects: a class of curves---timelike geodesics---on the one hand; and a class of smooth, symmetric tensor fields---namely those that are divergence-free and satisfy a certain energy condition---on the other.  Both of these classes of object are defined relative to a particular derivative operator, $\nabla$.  It is this structure that ultimately links the two classes: it is the tensor fields that are divergence-free relative to $\nabla$ that, in the small-body limit, must follow the geodesics of $\nabla$.  Indeed, this is the only structure relating the two classes of objects, in the sense that if one required in the statement of the Geroch-Jang theorem that each field $T^{ab}$ were conserved relative to some \emph{other} (metric) derivative operator, it would follow that in the small-body limit, matter would follow the geodesics of \emph{that} derivative operator instead.\footnote{I offer this claim without proof, though it essentially follows from the original proof of the Geroch-Jang theorem once one is attentive to subtleties concerning the energy condition, which makes reference to the metric, and not the derivative operator.  (Recall, however, footnote \ref{EC}: the work done by the energy condition is to define a certain cone; this could be done without reference to any metric.)  This particular relationship emerges more transparently if one uses the methods of \citet{Geroch+Weatherall}.  One could also imagine generalizing further, to the case where $\nabla$ is not the Levi-Civita derivative operator of any metric.}  And so, the key to matter obeying the geodesic principle is that $T^{ab}$ be divergence-free relative to the Levi-Civita derivative operator associated with the spacetime metric.

Finally, note that nothing in the theorem requires us to suppose that $T^{ab}$ represents energy-momentum, in the sense of encoding anything about particular measurements that an observer might make.  What is essential to the theorem (and its interpretation) is just that there be some property or other represented by a symmetric, rank-2 tensor field that is supported precisely where matter is present, and which is divergence-free relative to a specified derivative operator whenever that matter is non-interacting.  Thus, we see that the important point in the Geroch-Jang theorem is not just that matter is ``conserved'' in some sense or other: it is that matter is associated with some property that is suitably adapted to the background spacetime structure in the sense of being divergence-free relative to a particular derivative operator.  That the energy-momentum tensor associated with matter happens to reliably have these features merely makes the significance of the result more transparent.

\section{Conservation in General Relativity}\label{GR}

I will now specialize to GR.\footnote{For background, see the references in note \ref{GRbooks}.}  Here one considers relativistic spacetimes $(M,g_{ab})$ with matter fields having energy-momentum density $T^{ab}$, as sketched above.  Now, though, one requires that the spacetime metric and its derivatives are dynamically related to the distribution of energy and momentum throughout space and time according to \emph{Einstein's equation}\begin{equation}G_{ab} = 8\pi T_{ab},\end{equation} where $G_{ab}$ is the Einstein tensor defined in Eq. \eqref{EinsteinTensor} and $T^{ab}$ is the total energy-momentum density in the universe, i.e., the sum of the energy-momentum tensors associated with all matter fields.\footnote{Here and throughout I work in units where $c=G=1$.}

As noted in the introduction, and as Brown emphasizes, Einstein's equation has an immediate consequence concerning conservation \citep[p. 161]{Brown}.  The left-hand side of Einstein's equation, the Einstein tensor $G_{ab}$, is a function of a metric $g_{ab}$ and its associated Ricci curvature tensor and curvature scalar.  And it happens to follow from a mathematical fact, known as \emph{Bianchi's identity}, that the divergence of this tensor, taken with respect to the Levi-Civita derivative operator determined by $g_{ab}$, vanishes identically.  This fact holds generally, of any Riemannian or semi-Riemannian manifold at all, and is not a special consequence of GR.  But if the left-hand side of Einstein's equation is necessarily divergence-free, then so too must be the right-hand side, i.e., the total energy-momentum density.  In this sense, that energy-momentum density is conserved follows from Einstein's equation.

One has to be careful about this inference, however---particularly in the context of the Geroch-Jang theorem.  For one, as noted in the introduction, this inference concerns only matter that appears as a source in Einstein's equation.  Of course, insofar as one accepts GR, one should expect that all matter present in the actual universe contributes to the total energy-momentum tensor on the right-hand side of Einstein's equation.  But the geodesic principle is an idealization concerning matter that is not only physically small, in the sense of occupying vanishing spatial volume, but also of vanishing mass, such that its contribution to Einstein's equation can be neglected.\footnote{This point is emphasized in particular by \citet{Tamir}, though I do not think it is controversial.  Extended, massive matter will at best approximate geodesic motion.  Note that even the Ehlers-Geroch theorem ultimately concerns the motion of bodies only in the limit where their contribution to Einstein's equation becomes arbitrarily small; see also \citet{Wald+Gralla}, who derive leading order corrections to this limit.}  As noted in the previous section, this is reflected in the Geroch-Jang theorem by the fact that one works with a curve $\gamma$ in a fixed background spacetime, and one does not assume that the $T^{ab}$ fields associated with neighborhoods of $\gamma$ satisfy Einstein's equation for that fixed background spacetime.  Indeed, doing so would substantially limit the significance of the theorem, insofar as generic spacetimes are not solutions to Einstein's equation for matter with spatially compact support.

There is another concern about the inference.  Even if one focuses attention on source matter, Einstein's equation tells us only about the divergence of the \emph{total} energy-momentum content of the universe.  But the geodesic principle is supposed to apply more generally: it does not concern only matter that is propagating in regions were no other matter is present (i.e., where the energy-momentum associated with the small body is the total energy-momentum present); it is supposed to apply to all \emph{free} matter, i.e., any matter that happens not to be interacting with its environment, even when there is other matter present in that environment.  (Think, for instance, of neutrinos or dark matter passing through the earth.  One should expect these to approximately follow timelike geodesics, even though there is other matter present, because that matter is transparent to them.)  But we cannot infer from Einstein's equation that the energy-momentum associated with such matter is divergence-free.  This observation indicates that in the background of the interpretation of the Geroch-Jang theorem and similar results is a more general framework for thinking about influence and interaction between different matter fields in a region of spacetime, where interaction is signalled by exchange of energy-momentum.  This exchange is represented by a failure of the energy-momentum tensors associated with two or more matter fields to be individually divergence-free, even though their sum is divergence-free.  This framework for thinking about interaction (and, conversely, non-interaction) is more general than GR.

All that said, Einstein's equation does establish one relationship: it tells us with respect to what derivative operator the total energy-momentum tensor is conserved.  And in the special case where a small body is the only matter present in a region, and we suppose that it is taken as source matter, then Einstein's equation does imply that there is a property of that body represented by a symmetric rank-2 tensor and conserved relative to the Levi-Civita derivative operator.  Perhaps this does not directly imply that all non-interacting matter is conserved relative to this derivative operator, but it is a strong link between the dynamical properties of matter and the geometry of spacetime.

Still, one might wonder if this is the \emph{only} such link, or even the strongest one.  In fact, as I will argue presently, it is not the only link.  To see this, one must first address a prior question: given some variety of matter, how are we supposed to associate an energy-momentum tensor with it in the first place?  In other words, suppose you postulate some form of matter---some new object represented by a field, $\Psi^X$, say, where the $X$ indicates arbitrary index structure, satisfying some system of differential equations on spacetime.\footnote{We assume here that these indices are tangent indices.  One could extend this argument to matter represented by sections of a more general vector bundle, but doing so introduces further complications.}  Presumably there will be some energy-momentum tensor associated with this matter.  One might expect, based on experience with other matter fields, that this energy-momentum tensor will be some algebraic function of $\Psi^X$, and that it will have certain properties, such as that it will be non-vanishing in any open region were the field $\Psi^X$ is non-vanishing. But how are you to actually determine what $T^{ab}$ is?

In general, there is no strict procedure, even if you know the dynamics of the matter field.  \citet{Hawking+Ellis} put the point nicely, as follows:
\begin{quote}\singlespacing
The conditions (i) and (ii) of postulate (b) [that (i) total energy-momentum is non-vanishing on an open set iff the matter field vanishes there, and (ii) that total energy-momentum is divergence-free] do not tell one how to construct the energy-momentum tensor for a given set of fields, or whether it is unique.  In practice one relies heavily on one's intuitive knowledge of what energy and momentum are.  (p. 64)
\end{quote}
Insofar as the procedure relies on intuition and rules of thumb, the fact that the energy-momentum tensors we associate with matter---whether newly postulated matter, or even familiar matter fields such as electromagnetic fields---happen to be divergence-free when non-interacting may look arbitrary, unmotivated, or conventional.  From this perspective, the insight offered by Einstein's equation, that whatever else is the case total energy-momentum \emph{must} be conserved for Einstein's equation to make sense, looks particularly valuable.

But, Hawking and Ellis continue, there is an important and broad exception to the general rule just stated.  In the special case where the differential equations governing $\Psi^X$ may be derived by extremizing an action associated with a Lagrangian density, there is a ``definite and unique formula for the energy-momentum tensor'' associated with that matter (p. 64).  This formula for the tensor is forced upon us as the term that appears on the right-hand side of Einstein's equation, now understood as the Euler-Lagrange equation associated with extremizing the joint Einstein-Hilbert-matter action with respect to the spacetime metric.  (The Einstein-Hilbert action is the action of GR.)  Once we have this formula, however, we are in a position to investigate the properties associated with this energy-momentum tensor.  And as we will see, one can show that any energy-momentum tensor that arises in this way, in the absence of interactions with other matter---understood, now, as ``interaction terms'' in a joint Lagrangian density---must be divergence-free with respect to the covariant derivative operator.

The argument proceeds as follows.\footnote{The treatment here most closely follows that of \citet[pp. 64--67]{Hawking+Ellis}, with some elements borrowed from \citet[Appendix E.1]{Wald}.  See also \citet{TrautmanCLGR} and \citet[Appendix A]{Brown}; \citet{Holman} explains the relationship between this argument and Noether's theorems.  Of course, for many purposes it is more natural to think of a Lagrangian density as a scalar field on some appropriate jet bundle over spacetime, as in, for instance, \citet{Bleecker} or \citet{Saunders}.  But the extra generality of that approach is not necessary here, and so I prefer to work in a setting familiar from standard texts.  Note, too, that when I use the term ``density'', I mean only ``quantity I expect to integrate'', though there would be no contentful change if one interpreted the term instead in the sense of ``quantity that may be integrated even in the absence of orientability''.  Still, for simplicity, I will suppose that the manifold on which we are working is orientable.}  Let us begin by supposing that we have, associated with the field $\Psi^X$, a Lagrangian density, which we understand as a scalar field $\mathcal{L}[\Psi^X, g_{ab}]$ on spacetime whose value at each point is given by some fixed algebraic function, independent of the point, of the metric $g_{ab}$ at that point, $\Psi^X$ at that point, and the covariant derivatives of $\Psi^X$ at that point, up to some finite order.  This Lagrangian density may be integrated over compact regions $C$ of spacetime to define an \emph{action},
\begin{equation}
I_C[\Psi^X,g_{ab}] = \int_C \mathcal{L}(\Psi^X,g_{ab})dV,
\end{equation}
where $dV$ is a volume element determined by $g_{ab}$.  This action is a functional on field configurations (and their derivatives), i.e., a map from field configuration on $M$ to real numbers. From this action, we derive equations of motion for $\Psi^X$ by requiring that this action functional be functionally differentiable and stationary under variations of the field $\Psi^X$ that vanish outside of the interior of $C$, for all compact regions $C$.

More precisely, by a \emph{variation of the field} $\Psi^X$ about some value $\Psi^X_0$, I mean a one parameter family $s\mapsto\Psi^X(s)$, for $s$ defined on some open interval $(-\epsilon,\epsilon)\subseteq \mathbb{R}$, with the properties that: (1) $\Psi^X(0)=\Psi^X_0$ at each point $p$; (2) for all $s\in (-\epsilon,\epsilon)$, $\Psi^X(s) = \Psi^X_0$ outside of the interior of $C$; and (3) the family is differentiable with respect to $s$ at $s=0$.  Given such a variation, the derivative $\left(d\Psi^X/ds\right)_{|s=0}$ is denoted by $\delta \Psi^X_{|\Psi^X_0}$.  We will say that a functional $S[\Psi^X]$ is \emph{functionally differentiable} at $\Psi^X_0$ if there exists a tensor field, denoted $\left(\frac{\delta S}{\delta \Psi^X}\right)_{|\Psi^X_0}$, such that for all variations,
\[
\left(\frac{dS}{ds}\right)_{|s=0} = \int_M \left(\frac{\delta S}{\delta \Psi^X}\right)_{|\Psi^X_0} \delta \Psi^X_{|\Psi^X_0}.
\]
A functional is \emph{functionally differentiable} if it is functionally differentiable at all field values $\Psi^X_0$, in which case we define $\frac{\delta S}{\delta \Psi^X}$ as the functional derivative of $S$, where $\frac{\delta S}{\delta \Psi^X}$ is a map from field configurations to tensor fields whose evaluation at each field configuration $\Psi^X_0$ yields $\left(\frac{\delta S}{\delta \Psi^X}\right)_{|\Psi^X_0}$.  Note that the actions associated with known matter fields are all functionally differentiable.

An action $I_C[\Psi^X,g_{ab}]$ is said to be \emph{stationary} under variations of the field configuration $\Psi^X$ at a particular configuration $\Psi^X_0$ if $\left(\frac{\delta I_C}{\delta \Psi^X}\right)_{|\Psi^X_0}=0$.  This is known to occur at $\Psi^X_0$ if and only if $\Psi^X_0$ is a solution to the \emph{Euler-Lagrange equations}, which are differential equations on $M$ with $\Psi^X$ as the independent variable.  It is these Euler-Lagrange equations that are taken to be the field equations governing $\Psi^X$, and they are understood to be derived from the action in the sense that their solutions are values of $\Psi^X$ for which the action is stationary.  Note that since the Lagrangian density for $\Psi^X$ will in general depend on the metric $g_{ab}$, so too will the Euler-Lagrange equations.  Thus, a solution to the equations will only be a solution for some given metric field, $g^0_{ab}$.  To emphasize this important fact, I will write $\left(\frac{\delta I_C}{\delta \Psi^X}\right)_{|\Psi^X_0,g^0_{ab}}$ in what follows to indicate that any variation in $\Psi^X$ depends on the metric.

Of course, my purpose here is not to systematically develop Lagrangian field theory.  Rather, it is just to lay sufficient groundwork to make the following observation.  The functional $I_C$ is defined as an integral of some scalar field $\mathcal{L}$, which in turn is a function of tensor fields on $M$.  This means that if $\chi:M\rightarrow M$ is a diffeomorphism that acts as the identity outside of $C$, it must be the case that $I_C[\Psi^X,g_{ab}]=\int_C \mathcal{L}(\Psi^X,g_{ab})dV = \int_C \chi_*(\mathcal{L}(\Psi^X,g_{ab})dV) = \int_C \mathcal{L}(\chi_*(\Psi^X),\chi_*(g_{ab}))\chi_*(dV)) =  I_C[\chi_*(\Psi^X),\chi_*(g_{ab})]$, where the second equality follows from the general property that integrals are invariant under the pushforward map, and the third equality follow from the fact that $\mathcal{L}$ is a functional of fields and their derivatives at a point, independent of the point.\footnote{\label{covariance}This fact about actions is sometimes described as a ``principle of general covariance''. But this ``principle'' is just shorthand for the assertion that $I_C[\Psi^X,g_{ab}]=I_C[\chi_*(\Psi^X),\chi_*(g_{ab})]$ for any diffeomorphism of the form discussed above; it is not some additional principal above and beyond the facts that the fields under consideration are tensor fields and the Lagrangian density has a fixed functional form.  In particular, it is not ``imposed'' at this stage of the argument.  Of course, the fact that the dynamics of matter \emph{are} independent of position is a substantive physical fact, though it is not clear that it has anything to do with coordinate independence or general covariance (whatever that might mean), since any position-dependent dynamics could equally well be expressed in a coordinate-independent manner.}

But now suppose that one has some smooth vector field $w^a$, vanishing outside of $C$, and let $s\mapsto \chi_s$ be the one-parameter family of diffeomorphisms generated by $w^a$, for $s$ in some interval $(-\epsilon,\epsilon)$.  Then $s\mapsto \chi_s^*(\Psi^X_0)$ defines a variation of the field $\Psi^X$ about the configuration $\Psi^X_0$ and $s\mapsto \chi_s^*(g^0_{ab})$ defines a variation of $g_{ab}$ about the configuration $g^0_{ab}$.  It follows from the invariance property of $I$ just derived that:
\begin{equation}
0=\left(\frac{d I_C}{d s}\right)_{|s=0}.
\end{equation}
The functional integrability of $I_C$ then allows us to infer that,
\begin{equation}
0 = \int_C \left(\frac{\delta I_C}{\delta g_{ab}}\right)_{|\Psi^X_0,g^0_{ab}}\left(\delta g_{ab}\right)_{|g^0_{ab}} + \int_C \left(\frac{\delta I_C}{\delta \Psi^X}\right)_{\Psi^X_0,g^0_{ab}}\left(\delta\Psi^X\right)_{|\Psi_0^X},
\end{equation}
where we get two terms because we are simultaneously varying both $\Psi^X$ and $g_{ab}$.

Suppose, now, that $\Psi^X_0$ is a solution to its equations of motion, i.e., it is a stationary point for the action, evaluated at the metric $g^0_{ab}$.  Then $\left(\frac{\delta I_C}{\delta \Psi^X}\right)_{|\Psi^X_0,g^0_{ab}}=\mathbf{0}$.  It follows that
\begin{equation}
0 = \int_C \left(\frac{\delta I_C}{\delta g_{ab}}\right)_{|\Psi^X_0,g^0_{ab}}\left(\delta g_{ab}\right)_{|g^0_{ab}}.
\end{equation}
But recall that $\left(\delta g_{ab}]\right)_{|g^0_{ab}}$ was defined as $\left(dg_{ab}/ds\right)_{|s=0}$, which in this case is $(d\chi_s(g_{ab}^0)/ds)_{|s=0}=\pounds_w g^0_{ab} = 2\nabla^0_{(a}w_{b)}$, where $\nabla^0$ is the Levi-Civita derivative operator associated with $g^0_{ab}$, because the variations were determined by a one-parameter family of diffeomorphisms generated by the vector field $w^a$.  It follows that
\begin{equation}\label{lastStep}
0 = \int_C \left(\frac{\delta I_C}{\delta g_{ab}}\right)_{|\Psi^X_0,g^0_{ab}}\nabla^0_{(a}w_{b)} = \int_C T^{ab}\nabla^0_{(a}w_{b)}dV = -\int_C \left(\nabla^0_aT^{ab}\right)w_{b}dV,
\end{equation}
where we have defined a field $T^{ab}$ as the unique (symmetric) tensor field with the property that $T^{ab}dV = \left(\frac{\delta I_C}{\delta g_{ab}}\right)_{|\Psi^X_0,g^0_{ab}}$,\footnote{Here $dV$ is the volume element determined by $g_{ab}^0$.} and in the final equality we have integrated by parts, neglecting the term $\int_C\nabla^0_a(T^{ab}w_{b})dV$ because $w^a$ was defined to vanish on the boundary of $C$, which implies, by Stokes' theorem, $\int_C\nabla^0_a(T^{ab}w_{b})dV=0$.  Now, Eq. \eqref{lastStep} necessarily holds for all compact regions $C$ and all vector fields $w^a$ supported on $C$.  This can only happen if $\nabla_a^0 T^{ab} = \mathbf{0}$ everywhere.

Let us step back from the details of the calculations, now, and consider what has been established.  We have just seen, based on a simple invariance property of integrals under the pushforward map that any field configuration $\Psi^X_0$ that extremizes an action $I_C[\Psi^X,g_{ab}]$ for some metric $g_{ab}^0$ is associated with a symmetric rank-2 tensor field $T^{ab}$, defined by $T^{ab} dV = \left(\frac{\delta I_C}{\delta g_{ab}}\right)_{|\Psi^X_0,g^0_{ab}}$, that is divergence-free with respect to the Levi-Civita derivative operator compatible with $g^0_{ab}$.  Moreover, this quantity will be non-vanishing wherever $\Psi^X_0$ is non-vanishing, as long as $\mathcal{L}$ vanishes only where $\Psi^X_0$ does and contains terms that have non-trivial dependence on $g^0_{ab}$, as is the case for all Lagrangians encountered in GR.  Indeed, it turns out that this tensor field $T^{ab}$ is, up to a constant multiple, the energy-momentum tensor associated with $\Psi^X_0$, in the sense described above, i.e., that this is the term that would appear on the right-hand side of Einstein's equation, derived by extremizing an appropriate action.  But for present purposes, this is not essential.  All that matters is that there is some tensor quantity associated the matter field $\Psi^X_0$ that is symmetric, rank-2, vanishing on open regions only if $\Psi^X_0$ vanishes there, and divergence-free relative to the spacetime derivative operator.

Thus we have found that the existence of a divergence-free tensor $T^{ab}$ field associated with the matter field $\Psi^X$ may be understood as a consequence of the equations of motion of $\Psi^X$: when those equations hold, in the absence of interactions, there is a divergence-free quantity associated with the field, where the divergence is taken with respect to the covariant derivative operator.  As promised, this argument provides a general, independent justification for the conservation condition that does not run through Einstein's equation---and for this reason, as we will see in subsequent sections, may be applied to theories other than GR.  (As I discuss below, it also provides an independent justification for our association of the conservation condition with non-interaction.)  It follows that one can apply the Geroch-Jang theorem to conclude that insofar as this kind of matter may be confined to propagate along a curve, and if the matter satisfies an energy condition, then that curve must be a timelike geodesic.\footnote{It is worth pointing out, however, that a lot of work here is done by the caveat ``insofar as this kind of matter may be confined to propagate along a curve''. In fact, the limiting conditions of the Geroch-Jang theorem are strikingly strong, and solutions to many hyperbolic systems, including solutions to Maxwell's equations, the Klein-Gordon equation, and the Dirac equation, can never satisfy those limiting conditions because wave equations generally do not have solutions that are supported only in arbitrary spatially compact neighborhoods of a curve.  (That is, they spread out over time.)  This important issue is addressed by \citet{Geroch+Weatherall}, however, and so the limitation may be mitigated.}  In other words, the sort of matter represented by $\Psi^X$, in the small body limit, must respect the inertial structure of GR.

It is worth reflecting on what, in this derivation, makes the conservation condition true---or, put differently in light of the discussion of the previous section, how is it that a particular Levi-Civita derivative operator appears in this derivation?  The answer has two parts, but both depend on a basic assumption, which is that in the absence of interactions with other matter, the Lagrangian density for a given matter field depends (only) on the matter field itself, its covariant derivatives, and a metric with which the covariant derivative operator is compatible.\footnote{It is \emph{not}, as some authors would suggest, a ``principle of general covariance'' that does the important work, here.  (Recall footnote \ref{covariance}.)  It is true that the argument requires that one is working with tensorial quantities, so that a diffeomorphism may be extended to act on the dynamical fields.  But there are many derivative operators available on a manifold, and ``general covariance'' alone simply cannot pick out relative to which of them the field $T^{ab}$ is to be divergence-free.  This derivative operator---the link to the particular inertial trajectories---is selected by other considerations in the argument.}  It is for this reason that a solution to the equations of motion for $\Psi^X$ on a spacetime $(M,g_{ab})$, is a solution only relative to the metric $g_{ab}$.  This is the first part of the answer: it is the covariant derivative operator $\nabla$ determined by the metric relative to which $\Psi^X$ is a solution that appears in the conservation principle we end up with.

The second part of the answer is closely related: it is the same fact about the Lagrangian density that allows us to draw the crucial inference that the action must be stationary under certain joint variations of the fields, as a consequence of the fact that integrals are invariant under pushforwards.  This is what leads to the conclusion that $\nabla_a T^{ab}=\mathbf{0}$.  And so, ultimately the reason that solutions to the equations of motion for a matter field must be associated with a symmetric, rank-2, covariantly divergence-free tensor $T^{ab}$ is that the spacetime metric plays a certain role in the equations of motion for $\Psi^X$.  Or perhaps better: one can find a quantity $T^{ab}$ that is divergence-free relative to $\nabla$ because $\nabla$ is the unique torsion-free derivative operator determined by the metric that appears in the equations of motion for $\Psi^X$.

I noted above that one reason to take the field $T^{ab}$ derived in this way as an energy-momentum tensor is that if one were to extremize the joint Einstein-Hilbert-matter action with respect to the metric, it would turn out that a field proportional to $T^{ab}$ would appear in the resulting Euler-Lagrange equations in the place that the energy-momentum tensor should.  (Another reason to take it as an energy-momentum tensor is that the tensor derived in this way has many properties we usually associate with energy and momentum.)  So one might worry that this means that the whole argument just presented really amounts to nothing more than the original observation that whatever field appears opposite the Einstein tensor must be divergence-free.

I think this worry is too fast.  The main reason is that the argument just presented goes through even without assuming that the metric appearing in $T^{ab}$, or the dynamics of $\Psi^X$, is a solution to Einstein's equation for that particular source field.  (In other words, the Einstein-Hilbert action plays no role.)  What matters is \emph{only} that a metric play a certain role in the dynamics of the matter field, and not that the metric additionally bear the relation to that matter that is captured by Einstein's equation.  Indeed, as \citet[p. 456]{Wald} points out, one can run the same argument above with just the Einstein-Hilbert action, instead of a matter action, and conclude on its basis that the Einstein tensor must be conserved!  Thus, the (contracted) Bianchi identity may itself be viewed as a consequence of the role that the metric plays in the Einstein-Hilbert action.  From this perspective, the argument that conservation follows from Einstein's equation is itself just a special case of the more general argument just presented.

This last point may raise another concern, however. The arguments of this section have relied on variational methods, in such a way that one might get the impression that I am claiming that variational methods are somehow more fundamental than other ways of presenting the dynamics of fields, or that having a Lagrangian formulation is essential for a theory to be physically interesting.  But I do not mean to endorse this attitude towards action principles at all.

As I see it, Lagrangian methods are just one useful and general way of analyzing the properties of systems of coupled differential equations.  In this respect, they are not more or less fundamental than other means of discussing the same equations, such as by using methods from functional analysis or Hamiltonian methods.  Variational methods have some advantages, as I will discuss presently, but also some disadvantages, including that not all systems of equations can be derived from a variational principle.

In the present case, using Lagrangian methods has two main advantages.  One is that it allows one to discuss the dynamical properties of many different (thought perhaps not all possible) fields at once, to show how under very general circumstances one can find a tensor $T^{ab}$ with the properties necessary to run the Geroch-Jang theorem, without invoking Einstein's equation.  It also allows one to pinpoint, in a general setting, what features of the dynamics of physical fields are important for establishing the existence of such a field $T^{ab}$.  But this is ultimately a pragmatic virtue.  And it is certainly the case that one could begin, say, just with a system of equations for a field (such as the source-free Maxwell equations) and produce, through intuition, brute force, or experience with the energy and momentum properties of that field, a tensor $T^{ab}$ that one can then show is divergence-free by virtue of the fields in question satisfying their equations of motion.  As a way of establishing the conservation condition, this approach is no less legitimate or ``deep'' than the one I pursue here.  But it is difficult to do in an abstract or sufficiently general way.

The second advantage to using variational methods in the present argument is that it allows one to clearly say when a field is ``free'' or ``non-interacting''.  This occurs just when the action for the field may be written as above, i.e., with no fields appearing except the field whose dynamics one is studying and the metric.  (If other fields appear in the action, then one must run the argument for the full action for all such fields to derive the result, in which case the field $T^{ab}$ that is conserved will include all of the interacting fields.)   And so it is that one can show that non-interacting fields will satisfy the conservation condition even if there happen to be other fields present in the same region.  It is this fact that we needed in order to get the interpretation we wanted from the Geroch-Jang theorem, that all free small bodies (approximately) follow timelike geodesics.  Again, though, one could also establish this by assuming that there are no source terms in the differential equations governing the fields in question, and so it is not as if an action principle is somehow essential to capture the present notion of ``non-interaction''.



\section{Conservation in Other Relativistic Theories}\label{alternatives}

In the last section, I began by specializing to GR, in order to discuss the relationship between Einstein's equation and conservation.  But the derivation of the conservation condition that I ultimately defended, like the Geroch-Jang theorem itself, did not depend on GR, in the sense that neither Einstein's equation nor the Einstein-Hilbert action played any role.  What mattered was only that there was some metric appearing in the matter Lagrangian density, and that the Lagrangian density depended only on that metric, the fields themselves, and their covariant derivatives.  One important consequence of this independence from Einstein's equation is that the same argument may be run in other relativistic theories of gravitation in which the analog of Einstein's equation does not imply that there is a quantity $T^{ab}$ associated with matter that is divergence-free.  Thus, even in these alternative theories, one can invoke the Geroch-Jang theorem, and the argument of the last section, to conclude that inertial motion has the status of a theorem, not a postulate.  I will now briefly illustrate this point with some examples.

One simple such example is known as unimodular gravity, or sometimes trace-free Einstein gravity \citep{Anderson+Finkelstein, Unruh, Ellis+etal}.  Unimodular gravity is a variation on GR in which one moves from Einstein's equation to a ``trace-free'' equation,\footnote{This equation is ``trace-free'' because it results by equating the trace-free parts of the Einstein tensor and (a constant multiple of) the energy-momentum tensor.  One can confirm that both $R_{ab}-\frac{1}{4} R g_{ab}$ and $T_{ab} - \frac{1}{4}Tg_{ab}$ have vanishing trace.}
\begin{equation}\label{TFEE}
R_{ab} - \frac{1}{4}Rg_{ab} = 8\pi\left(T_{ab} - \frac{1}{4} Tg_{ab}\right),
\end{equation}
where $T=T^a{}_a$ is the trace of the energy-momentum tensor.

One motivation for moving to this equation is that terms of the form $\Lambda g_{ab}$, which have no trace-free part, do not appear, and so physical sources---i.e., contributions to the total energy-momentum of the universe---that might mimic a cosmological constant, such as those sometimes associated with quantum vacuum energies in semi-classical gravitation, do not contribute to the right-hand side of the equation.  Thus, it is argued, unimodular gravity solves the so-called ``cosmological constant problem'', at least on one understanding of what that problem is supposed to be.\footnote{For further discussion of unimodular gravity and the cosmological constant problem, see \citet{Finkelstein+etal}, \citet{Earman1}, \citet{Ellis+etal}, \citet{Ellis}, and \citet{Schneider}.}  The trace-free Einstein equation is of interest for present purposes because it does not imply that $T^{ab}$ is divergence-free.  However, one can show that if we assume, in addition to Eq. \eqref{TFEE}, that $\nabla_bT^{ab}=\mathbf{0}$, then one can recover Einstein's equation, up to a term $\Lambda_0 g_{ab}$, where $\Lambda_0$ is an unspecified constant of integration.  Thus, one may understand unimodular gravity, in the presence of the conservation condition, to be equivalent to GR modulo a cosmological constant term.

So how are we to understand the conservation condition in this theory?  One might first think that insofar as it is independent of Eq. \eqref{TFEE}, the conservation condition is now an additional brute postulate, which, together with Eq. \eqref{TFEE}, implies Einstein's equation.  But this is too fast.  The reason is that the differential equations governing matter in unimodular gravity are supposed to be precisely the same as in GR.  In other words, the same Lagrangian densities, with the same dependencies on the metric $g_{ab}$ and the fields and their derivatives, would describe the dynamics of matter in unimodular gravity.  And as we have just seen, it is a consequence of the dynamics governing these matter fields that they are associated with a divergence-free quantity $T^{ab}$. In other words, the modifications to Einstein's equation involved in moving from GR to unimodular gravity do not affect the conservation condition, at least insofar as one understands the conservation condition to be underwritten by the dynamics of matter.  One can conclude, then, that the conservation condition is not an independent postulate and that the relationship between unimodular gravity and GR is even tighter than sometimes suggested.  More importantly for present purposes, one can invoke the Geroch-Jang theorem just as in GR and conclude that inertial motion in unimodular gravity should be construed as a theorem, with assumptions of just the same strength.\footnote{In particular, note that since the matter dynamics in unimodular gravity are precisely those of GR, one should expect the energy condition to be satisfied under just the same circumstances.}

A second example is known as Tensor-Vector-Scalar gravity, or TeVeS for short.  TeVeS was introduced by \citet{Bekenstein} as a relativistic generalization of so-called Modified Newtonian dynamics, or MOND, a theory developed by \citet{Milgrom} to account for astronomical phenomena that would otherwise be explained by dark matter (without positing dark matter).  It is a theory in which one posits, in addition to the metric $g_{ab}$, two further fields that govern ``gravitational'' interactions: a vector field $u^a$, required to be of unit length with respect to $g_{ab}$, and a scalar field, $\phi$.  The Lagrangian for these fields may be written as a sum of the Einstein-Hilbert Lagrangian and two other terms, governing the dynamics of $u^a$ and $\phi$, respectively.  The full details of the dynamics do not matter for present purposes; it suffices to observe that the analogue to Einstein's equation that one derives by varying the total action with respect to the metric $g_{ab}$ does \emph{not} imply that the analogue to $T^{ab}$ in those equations is divergence-free with respect to $\nabla$, the derivative operator compatible with $g_{ab}$.

Indeed, in general the energy-momentum tensor is \emph{not} divergence-free with respect to $\nabla$ in TeVeS.  The reason is that in addition to the fields just described, one also defines a field known as the \emph{physical metric}:
\begin{equation}\label{physMetric}
\tilde{g}^{ab} = e^{2\phi} g^{ab} - 2u^a u^b \sinh(2\phi),
\end{equation}
with inverse $\tilde{g}_{ab}$ and associated Levi-Civita derivative operator $\tilde{\nabla}$.  One then posits that it is \emph{this} field that plays the role of the metric in the dynamics of matter.  In other words, given a matter field $\Psi^X$, one generically expects the (non-interacting) Lagrangian density associated with $\Psi^X$ to be a fixed algebraic function, depending only on $\tilde{g}_{ab}$, $\Psi^X$, and the covariant derivatives of $\Psi^X$ with respect to $\tilde{\nabla}$; likewise the matter action is defined by integrating with respect to $d\tilde{V}$, the volume element determined by $\tilde{g}_{ab}$.

It follows that if one were to run the argument of the previous section, one would conclude that it is $\tilde{T}^{ab}$, the unique field such that $\tilde{T}^{ab}d\tilde{V} = \left(\frac{\delta I_C}{\delta \tilde{g}_{ab}}\right)_{|\Psi^X_0,g^0_{ab}}$, that is divergence-free when the equations of motion for $\Psi^X$ are satisfied, and more, that it is divergence-free with respect to $\tilde{\nabla}$, not $\nabla$.  Likewise, one should expect the Geroch-Jang theorem to apply---now with the consequence that in the small-body limit, non-interacting matter should follow $\tilde{g}_{ab}-$timelike geodesics of $\tilde{\nabla}$.  Thus, in TeVeS as in GR, one can understand inertial motion to be a theorem, not a postulate.  And that theorem tells us that the geodesics probed by small bodies are those of $\tilde{g}_{ab}$, not $g_{ab}$.  Insofar as conservation is taken to follow from the matter dynamics, then, one can take TeVeS to explain inertial motion just as well as GR, where now the explanation is that the matter dynamics are such that the effective metric appearing in those dynamics is $\tilde{g}_{ab}$.  One might even argue that inertial motion is more strongly in need of explanation in TeVeS than in GR, insofar as there are multiple fields in the theory that may be taken to define candidate classes of privileged curves.

The line I have just adopted is, in many ways, compatible with Brown's own remarks about TeVeS \citep[\S 9.5.2]{Brown}.  As he observes,
\begin{quote}\singlespacing The metric field that is surveyed by rods and clocks, whose conformal structure is traced by light rays and whose geodesics correspond to the motion of free bodies is clearly not [$g_{ab}$], but the less `basic' [$\tilde{g}_{ab}$]. ... It is this field, and not the metric [$g_{ab}$] ..., which acquires chronometric significance, and it does so because of the postulated dynamics in the theory.  Right or wrong, the theory reminds us that the operational significance of a non-singular second-rank tensor field, and its geometric meaning, if any, depends on whether it is `delineated by matter dynamics'. (p. 175)\end{quote}
On this much I have no objection: the metric $\tilde{g}_{ab}$ determines inertial structure in TeVeS precisely because of the role it plays in the matter dynamics.

But the link between inertia and dynamics that runs through conservation is secured only by the argument given in the previous section, and not by any argument involving the TeVeS analogs to Einstein's equation.  Brown acknowledges as much when he writes, in a footnote, that:
\begin{quote}\singlespacing The vanishing of the covariant divergence of [$\tilde{T}^{ab}$]---with respect to [$\tilde{g}_{ab}$]---is required if the geodesic principle of free motion and the local validity of Maxwell's equations are to hold, and this requirement is not a consequence of the field equations [for TeVeS] alone.  It is a remarkable feature of the other field equations that they conspire, together with the [TeVeS equations], to ensure the validity of the `conservation' requirement. (p. 175, fn. 66)\end{quote}
The first sentence is unobjectionable.  But the second, I would suggest, is at best misleading, because in the end the TeVeS equations themselves---the analog to Einstein's equation in particular---play no role at all in the fact that $\tilde{T}^{ab}$ is divergence-free, just as, in the argument in the previous section, Einstein's equation played no role.  More, there is nothing ``remarkable'' (or ``conspiratorial'') about the fact that $\tilde{T}^{ab}$ is divergence-free with respect to $\tilde{\nabla}$.  It is a basic consequence of the fact that it is $\tilde{g}_{ab}$ that plays the requisite role in the equations governing matter.  In other words, if Brown wants to accept that TeVeS, by virtue of all of the field equations taken together, ensures that $\tilde{T}^{ab}$ is divergence-free, it seems he needs to accept the cogency of the sort of argument given in the previous section.  But if that argument is cogent in TeVeS and unimodular gravity (assuming Brown accepts that it is), it is hard to deny its cogency in GR.

This discussion leads to a final remark, concerning the status of the conservation condition in SR.  Brown emphasizes throughout \emph{Physical Relativity} that GR is to be distinguished from SR with respect to the status of inertial motion, precisely because it is a theorem in GR and not SR.  But both the Geroch-Jang theorem and the argument of the previous section apply equally well in SR as in GR, insofar as it is the Minkowski metric that plays the required role in the dynamics for matter.  And so, it would seem, both inertial motion and conservation of matter should be viewed as theorems of SR just as much as GR (or unimodular gravity, or TeVeS), with much the same status in both cases.  To resist this conclusion, it would seem that the only line available to Brown would be to reject the arguments in the previous section as somehow illegitimate, or insufficiently explanatory, to justify claiming that SR explains inertial motion.  I do not see the attraction of this line of argument, especially for Brown, who also argues that it is precisely the equations governing matter that justify the operational significance of geometrical structures.  But whatever else is the case, if one were to adopt this line, it would seem to have the unpleasant consequence that TeVeS and unimodular gravity, like SR, cannot be said to explain inertial motion---even though Brown, at least, wants to take TeVeS as a paradigm example of the relationship between effective geometry and matter dynamics.

\section{Conservation in Newtonian Theories}\label{GNG}

I will now turn to Newtonian theories. The focus, here, will be on GNG, but just as what was said about GR in section \ref{GR} turned out to apply equally well in SR, so, too, will what I say about GNG apply to ordinary Newtonian gravitation.  The setting now is a \emph{classical spacetime}, which is a quadruple $(M,t_{a},h^{ab},\nabla)$, where $M$ is a smooth, four-dimensional manifold that is once again Hausdorff and second countable; $t_{a}$ is a smooth one-form on $M$;\footnote{Note that by working with a rank-1 field $t_a$ instead of a rank-2 field $t_{ab}$, I am assuming that spacetime is temporally orientable.} $h^{ab}$ is a smooth symmetric field on $M$ of signature $(0,1,1,1)$, satisfying $h^{ab}t_{b}=\mathbf{0}$; and $\nabla$ is a derivative operator on $M$ that is compatible with $t_{a}$ and $h^{ab}$ in the sense that $\nabla_a t_{b}=\mathbf{0}$ and $\nabla_a h^{bc}=\mathbf{0}$.\footnote{For background and details on classical spacetimes, including GNG, see \citet{Trautman} and (especially) \citet[Ch. 4]{MalamentGR}.  See also the references in footnote \label{GNGrefs}.}

As in GR and other relativistic spacetime theories, points of the manifold $M$ represent events in space and time. The fields $t_{a}$ and $h^{ab}$ may be understood as (degenerate) temporal and spatial metrics, respectively, as follows.  Given a vector $\xi^a$ at a point, one says that $\xi^a$ is \emph{timelike} if $t_{a}\xi^a\neq \mathbf{0}$.  Otherwise $\xi^a$ is \emph{spacelike}.  If $\xi^a$ is timelike, then it has \emph{temporal length} $|t_a\xi^a|$ and it is \emph{future-directed} if $t_a\xi^a>0$.  If $\xi^a$ is spacelike, then there always exists a (non-unique) covector $\sigma_a$ such that $\xi^a=h^{ab}\sigma_b$; we define the \emph{spatial length} of $\xi^a$ to be $\sqrt{h^{ab}\sigma_a\sigma_b}$, where it can be shown that this number is independent of the choice of $\sigma_a$.  It is important to emphasize that the temporal and spatial metrics do \emph{not} uniquely determine a compatible derivative operator $\nabla$: there are many derivative operators, including both flat and curved ones, that are compatible with the metrics, and so the derivative operator is a substantial additional piece of structure.  Given this derivative operator, one may define geodesics precisely as in a relativistic spacetime, and they again have an interpretation as unaccelerated curves.

In this context, one can prove the following analogue to the Geroch-Jang theorem.
\begin{thm} \emph{\textbf{ \citep{WeatherallMPNT}}}
\label{W}
Let $(M,t_{a},h^{ab},\nabla)$ be a classical spacetime satisfying $R^{ab}{}_{cd}=\mathbf{0}$.\footnote{\label{raising} Here and throughout this section we raise indices with $h^{ab}$, so that $R^{ab}{}_{cd} = h^{bn}R^a{}_{ncd}$.  Note, however, that $h^{ab}$ is not invertible, and so one cannot unambiguously lower and index that has been raised in this way.}  Let $\gamma:I\rightarrow M$ be a smooth imbedded curve.  Suppose that given any open subset $O$ of $M$ containing $\gamma[I]$, there exists a smooth symmetric field $T^{ab}$ with the following properties.
\begin{enumerate}
\item\label{mass} $T^{ab}$ satisfies the mass condition, i.e., whenever $T^{ab}\neq \mathbf{0}$, $T^{ab}t_at_b>0$;
\item\label{cons2} $T^{ab}$ satisfies the conservation condition, i.e., $\nabla_a T^{ab}=\mathbf{0}$;
\item\label{inside2} $\supp{T^{ab}}\subset O$; and
\item\label{non-vanishing2} there is at least one point in $O$ at which $T^{ab}\neq \mathbf{0}$.
\end{enumerate}
Then $\gamma$ is a timelike curve that can be reparametrized as a geodesic.
\end{thm}

The interpretation of this theorem is essentially the same as that of the Geroch-Jang theorem. Once again, we take matter to be generically represented by fields on $M$, where each of the various sorts of matter one encounters are associated with a smooth, symmetric rank-2 tensor, $T^{ab}$.  Now this tensor encodes the mass and momentum densities as determined by arbitrary observers.\footnote{For discussions of how $T^{ab}$ encodes this information, see especially \citet[\S4.1]{MalamentGR}; see also \citet{Duval+Kunzle}.}  Once again this tensor vanishes in an open region if and only if there is no matter present in that region.  Thus, the $T^{ab}$ fields appearing in the theorem, as in the Geroch-Jang theorem, characterize generic configurations of matter localized near a curve $\gamma$.

Also as in the Geroch-Jang theorem, this matter must satisfy two conditions.  The first is a condition that captures a sense in which $T^{ab}$ is ``positive'', now in the sense that mass density is always positive and there are no fields whose mass vanishes anywhere that the field is present.  And we have a conservation condition to capture the idea that the matter is free. As in the Geroch-Jang theorem, these conditions are supposed to hold of generic, physically reasonable, massive matter that is not interacting with its environment.  The only additional constraint in the present theorem is that we require that $R^{ab}{}_{cd}=\mathbf{0}$.  This curvature condition is one of two conditions necessary for the recovery of ordinary Newtonian gravitation from a model of GNG (the other is that $R^a{}_b{}^c{}_d = R^{c}{}_d{}^a{}_b$); one might understand both as criteria for a spacetime to be ``Newtonian''.\footnote{See \citet{Trautman} and \citet[\S 4.2]{MalamentGR} for discussions.}  Thus, the theorem may be interpreted just as in GR: in a suitably Newtonian spacetime, the only curves along which free, arbitrarily small ``bodies,'' i.e., spatially compact matter configurations, may propagate are timelike geodesics.  And so this theorem, too, captures the geodesic principle in GNG, in much the same way that the Geroch-Jang theorem captures it in GR.

Once again setting the energy (mass) condition aside, one may ask: what is the status of the conservation condition in GNG?  Note that there is no analogue to the first argument we considered in the context of GR, concerning Einstein's equation.  This is because the analogue to Einstein's equation in GNG, the \emph{geometrized Poisson equation}, states that
\begin{equation}
R_{ab} = 4\pi\rho t_at_b
\end{equation}
where $\rho=T^{mn}t_mt_n$ is the mass density associated with the total matter in the universe as determined by any (and hence, all) observer(s).  The mass-momentum tensor $T^{ab}$ does not appear in this equation in an uncontracted form, and so it is not clear how the equation could have anything to say about the divergence of that tensor; moreover, all indices in the equation are in lowered positions, and so to take a divergence one of them would need to be raised, i.e., an inner product would need to be taken.  The only object available to take inner products, however, is $h^{ab}$, and the right-hand side of the equation is orthogonal to $h^{ab}$ in both indices.  Hence raising indices with $h^{ab}$ would annihilate both sides of the equation.  So it seems that the geometrized Poisson equation cannot imply that $T^{ab}$ is divergence-free.

But there is another route to establish that $T^{ab}$ must be divergence-free---or at least, that there is some symmetric, rank-2 tensor field $T^{ab}$, divergence-free with respect to the covariant derivative operator defining the spacetime, associated with generic matter distributions in the absence of interaction.  The argument is due to \citet{Duval+Kunzle}.\footnote{\citet{Duval+KunzleSE} develop the argument in the context of a concrete example; see also \citet{Christian}, who presents the argument in somewhat more, or at least different, detail.}  As in GR, it makes use of variational methods, though the differences in geometrical setting between relativistic and classical spacetimes require some modifications.

Suppose we are again interested in a field $\Psi^X$, now in the context of a classical spacetime $(M,t_{a},h^{ab},\nabla)$.  Let us suppose again, as in the relativistic case, that we have, associated with the field $\Psi^X$, a Lagrangian density, which we now understand as a scalar field $\mathcal{L}(\Psi^X, t_{a},h^{ab},\nabla)$ on spacetime.  Once again this Lagrangian is required to be such that its value at each point is given by some fixed algebraic function, independent of the point.  Now, however, the Lagrangian density is allowed to depend not only on $\Psi^X$ and its covariant derivatives (relative to $\nabla$) at that point, but also on the spacetime structures $t_{a}$, $h^{ab}$, and $\nabla$, all of which would generically be expected to appear in the dynamics of a field on classical spacetime.  Once again we may integrate the Lagrangian density over a compact region $C$ of spacetime to define an action,
\begin{equation}
I_C[\Psi^X,t_{a},h^{ab},\nabla] = \int_C \mathcal{L}(\Psi^X, t_{a},h^{ab},\nabla)dV,
\end{equation}
where now $dV$ is the volume element determined by the classical metrics $t_{a}$ and $h^{ab}$.\footnote{For further details on this volume element, see  \citet{WeatherallMPNT}.}  We assume this action is functionally differentiable.

We would now like to run the same argument that we ran in the relativistic case.  The observation with which that argument began is unaffected by the new context, \emph{mutatis mutandis}: if $\chi:M\rightarrow M$ is a diffeomorphism that acts as the identity outside of $C$, it must be the case that $I_C[\Psi^X,t_{a},h^{ab},\nabla]=\int_C \mathcal{L}(\Psi^X,t_{a},h^{ab},\nabla)dV = \int_C \chi_*(\mathcal{L}(t_{a},h^{ab},\nabla)dV) = \int_C \mathcal{L}(\chi_*(\Psi^X),\chi_*(t_{a}),\chi_*(h^{ab}),\chi_*(\nabla)))\chi_*(dV)) =  I_C[\chi_*(\Psi^X),\chi_*(t_{a}),\chi_*(h^{ab}),\chi_*(\nabla)]$.  Once again, the essential step is the second equality, which follows from the same general property of integrals as in the relativistic case.

Following the same argument, then, suppose that we have a one parameter family $s\mapsto \chi_s$ of diffeomorphisms generated by some smooth vector field $w^a$, vanishing outside of $C$, for $s$ in some interval $(-\epsilon,\epsilon)$.  This one parameter family again defines variations of the fields about any fixed values $\Psi_0^X$, $t^0_{ab}$, $h_0^{ab}$, and $\nabla^0$.  Suppose that $\Psi_0^X$ happens to be a solution to its equations of motion, for $t^0_{ab}$, $h_0^{ab}$, and $\nabla^0$.  We then find, invoking the invariance property just established, functional integrability, and the fact that $\delta\Psi^X_{|\Psi^X_0}=\mathbf{0}$ that:\footnote{Note that in Eq. \eqref{GNGCal2} and in (most of) the remainder of the section, I have suppressed that all relevant expressions are to be evaluated at $s=0$, i.e., at $\Psi^X_0$, $t_a^0$, $h^{ab}_0$, and $\nabla^0$, to avoid cumbersome notation.}
\begin{equation}\label{GNGCal2}
0=\int_C Q^a\delta t_{a}dV + \int_C P_{ab}\delta h^{ab}dV + \int_C \frac{\delta I_C}{\delta \nabla}\delta \nabla,
\end{equation}
where we have defined $Q^a$ to be the unique tensor field such that $Q^adV = \frac{\delta I_C}{\delta t_a}$, and likewise $P_{ab}$ is the unique field such that $P_{ab}dV = \frac{\delta I_C}{\delta h^{ab}}$.

The variations of $t_{a}$ and $h^{ab}$ under this one parameter family of diffeomorphisms are easy to describe: they are, by definition, Lie derivatives with respect to $w^a$, so that $\left(\delta t_{a}\right)_{|t^0_{a}} = \pounds_w t^0_{a}= t_n\nabla^0_{a}w^n$ and $\left(\delta h^{ab}\right)_{|h_0^{ab}}=\pounds_w h_0^{ab} = -2\nabla^{0(a}w^{b)}$.  But variations of $\nabla$ are somewhat more difficult to describe.  Following \citet{Duval+Kunzle}, we characterize them as follows.  First, note that given \emph{any} unit timelike vector field $u^a$ on $M$, there exists a unique derivative operator $\overset{u}{\nabla}$ on $M$ that is compatible with $t^0_a$ and $h_0^{ab}$, and which is such that $u^n\overset{u}{\nabla}_n u^a = \mathbf{0}$ and $\overset{u}{\nabla}{}^{[a}u^{b]} = \mathbf{0}$.\footnote{Here $\nabla^{[a}u^{b]} = h^{n[a}_0\overset{u}{\nabla}_nu^{b]}$.  Recall fn. \ref{raising}.}  Meanwhile, given two derivative operators $\nabla$ and $\nabla'$ on $M$, both compatible with $t^0_a$ and $h^{ab}_0$, there always exists a two-form $\kappa_{ab}$, which is such that $\nabla' = (\nabla,2h_0^{an}t^0_{(b}\kappa_{c)n})$.  Thus, in particular, any derivative operator $\nabla$ compatible with $t^0_{a}$ and $h_0^{ab}$ is uniquely specified by a pair $(u^a,\kappa_{ab})$, where $u^a$ is a unit timelike vector $u^a$ and $\kappa_{ab}$ is a 2-form.  Finally, one can show that the 2-form appearing in this pair is closed if and only if $\nabla$ satisfies the ``Newtonian'' curvature condition, $R^a{}_b{}^c{}_d = R^c{}_d{}^a{}_b$; in what follows, we will limit attention to derivative operators satisfying this condition.  At least locally, then, there always exists a one form $A_a$ such that $\kappa_{ab}=\frac{1}{2}d_aA_b$, where $d_a$ is the exterior derivative.  So (again, locally), we may uniquely specify a derivative operator by a pair $(u^a,A_a)$.

Specifying a derivative operator in this way has the useful feature that if $\nabla^0$ is specified by the pair $(u^a,A_a)$, then for any diffeomorphism $\chi$, $\chi^*(\nabla^0)$ is specified by $(\chi^*(u^a),\chi^*(A_a))$.  More generally, any variation in $\nabla^0$ may be thought of as a suitable simultaneous variation of $u^a$ and $A_a$.  This means that we may re-express the final term in Eq. \eqref{GNGCal2} as,
\begin{equation}\label{GNGCal3}
\int_C \frac{\delta I_C}{\delta \nabla}\delta \nabla = \int_CK_a \delta u^adV + \int_CJ^a\delta A_adV,
\end{equation}
where $K_a$ is defined so that $K_adV = \frac{\delta I_C}{\delta u^a}$ and $J^a$ is such that $J^a dV = \frac{\delta I_C}{\delta A_a}$.  And since $u^a$ and $A_a$ are tensor fields, variations induced by the one parameter family of diffeomorphisms generated by $w^a$ may be expressed as Lie derivatives.

There is one final step.  The expression of a derivative operator in terms of the pair $(u^a,A_a)$ is not unique.  This means that there are some joint variations of $u^a$ and $A_a$ that do not change $\nabla^0$ (or $t_a^0$ or $h_0^{ab}$).  With some algebra, one can show that these are variations of $u^a$ and $A_a$ such that $t^0_a\delta u^a  = 0$, and which jointly satisfy the condition that $\delta A_a = \hat{h}^0_{an}\delta u^n + d_a \zeta$, where $\zeta$ is an arbitrary scalar field and $\hat{h}^0_{an}$ is the unique tensor field such that (1) $\hat{h}^0_{ab}h_0^{bc} = \delta_a{}^c - t^0_a u^c$ and (2) $\hat{h}^0_{ab}u^b=\mathbf{0}$ \citep[p. 358]{Duval+Kunzle}.  Observing that $\delta\nabla$ must vanish for variation of this sort, we find that:
\begin{equation}
0 = \int_C\left[(K_a + \hat{h}^0_{an}J^n)\delta u^a  + J^a d_a \zeta) \right]dV.
\end{equation}
Since this must hold for arbitrary variations $\delta u^a$ and for all smooth $\zeta$, we conclude that $\nabla_a J^a=0$ and that $K_a=-\hat{h}^0_{an}J^a + Xt_a$, where $X$ is an (unspecified) constant.

Combining Eqs. \eqref{GNGCal2} and \eqref{GNGCal3} with this last expression for $K_a$ in terms of $J^a$ and $X$ allows us to finally conclude that
\begin{align}\label{GNGlastStep}
0 &= \int_C \left(Q^a t^0_n\nabla^0_{a}w^n -2P_{ab}\nabla^{0(a}w^{b)} + J^a \pounds_w A_a + (-\hat{h}^0_{an}J^n + Xt^0_a)\pounds_w u^a\right)dV \notag\\
&= \int_C \left(-\nabla^0_b\tilde{T}^{b}{}_a + 2J^b \kappa_{ab} -\hat{h}^0_{mn}J^n\nabla^0_a u^m\right)w^a dV,
\end{align}
where we have, in the second equality, recalled that $\kappa_{ab}=d_aA_b$, integrated by parts, and combined terms to define a field $\tilde{T}^{a}{}_b$.  Observing that if Eq. \eqref{GNGlastStep} is to hold for all vector fields $w^a$ with support $C$, and for all compact sets (with sufficiently small interior), we conclude that the divergence of $\tilde{T}^a{}_b$ must be given by,
\begin{equation}\label{Ttilde}
\nabla^0_a\tilde{T}^{a}{}_b = 2J^a \kappa_{ba} -\hat{h}^0_{mn}J^n\nabla^0_a u^m
\end{equation}

The field $J^a$ is naturally interpreted as a mass-current density.  With this interpretation in mind, we impose the following condition: we suppose that $J^a$ is timelike and future-directed when it is non-vanishing.  This condition is precisely the mass condition used in the classical analog to the Geroch-Jang theorem.  (Note, however, that though the proposed interpretation motivates this condition, the condition may be imposed without comment on the interpretation of the fields.)  This assumption allows us to write $J^a$ at each point where it is non-vanishing as $J^a=\rho \eta^a$, where $\rho=J^at_a$ and $\eta^a$ is a unit timelike vector.  It also allows us to define a symmetric tensor $T^{ab}$ as follows:
\begin{equation}
T^{ab} = J^{a}\eta^{b} + \sigma^{ab},
\end{equation}
where $\sigma^{ab} = -\tilde{T}^{ab} + J^a\eta^b - J^au^b$ is symmetric and spacelike in both indices.  Using the known divergences of $J^a$ and $\tilde{T}^{a}{}_b$, we conclude that that $\nabla^0_b T^{ab}=\mathbf{0}$.  Thus, associated with generic matter fields in GNG is a symmetric rank-2 tensor $T^{ab}$ that is divergence-free relative to $\nabla^0$.

The upshot of this discussion is substantially the same as in the relativistic theories already discussed.  Once again, it follows from the dynamics of matter in a classical spacetime that, if those dynamics may be expressed via a Lagrangian density that depends only on the fields describing the matter and the classical spacetime structures $t_a$, $h^{ab}$, and $\nabla$, then there exists a symmetric rank-2 tensor field associated with the matter field that is divergence-free relative to the covariant derivative operator associated with the classical spacetime.  The reason for this is just the same as before: it is \emph{this} derivative operator, and compatible metrics, that appear in the matter equations of motion. And so, if one accepts the conclusion of the last two sections, then it would seem that the status of the conservation condition in Newtonian theories is very similar to in relativistic theories, including GR.  In light of the Newtonian analog to the Geroch-Jang theorem, then, it would seem that the status of inertial motion in all of these theories is essentially the same.  In all cases, the geodesic principle arises as the small-body limiting behavior of generic matter satisfying field equations in which the relevant geometric structures play a certain role, and no other such structures appear.

This is not to say that no differences can be identified between the Newtonian and relativistic cases.  There are some.  For instance, in the Newtonian case, and not the relativistic case, one requires a curvature condition, $R^a{}_b{}^c{}_d=R^c{}_d{}^a{}_b$, in order to derive the conservation condition.  If one drops this condition, one cannot conclude that $\kappa_{ab}$ is locally exact, and so one cannot find a vector field $A_a$ such that $\kappa_{ab} = d_aA_b$.  Thus one cannot necessarily define a vector field $J^a$ with the properties derived above; instead, one should expect to derive an antisymmetric rank-2 tensor.  It is possible that one could make an argument like the one above in this case as well, though it is not clear what the physical interest of doing so would be, insofar as we are interested in ``Newtonian'' theories.

It is also interesting to note that an energy condition appears to be necessary in the Newtonian case to arrive at the tensor $T^{ab}$ in the first place.  This suggests that the mass condition plays a deeper role in inertial motion in Newtonian theories than in GR (though of course some energy condition is ultimately necessary in both theories).  On the other hand, the mass condition is not very strong, and at least in the derivation of $T^{ab}$, its role is to rule out only pathological cases in which matter is not associated with any mass at all.\footnote{That said, see \citet{Dewar+Weatherall}: insofar as gravitation is associated with a mass-momentum tensor in Newtonian gravitation, it does not satisfy the mass condition.}  Note, too, that although the mass condition must hold, it still need not be the case that $J^a$ or $T^{ab}$ be interpreted as a mass-current density or mass-momentum density, respectively.  As long as there exist fields with the required positivity and conservation properties, the geodesic principle theorem goes through.  This last point leads to a final difference, which is that we have many fewer worked examples of matter fields in GNG that are characterized by an action principle.\footnote{The main example, already cited above, is that of a Schr\"odinger field in a classical spacetime, as studied by \citet{Duval+KunzleSE} and \citet{Christian}.  It would be interesting to develop a Lagrangian theory of perfect fluids in the classical context.}  This situation has the consequence that there are fewer examples in which we can observe that $T^{ab}$ as derived above is in fact the mass-momentum density associated with some physical matter described by a Lagrangian density.  But this worry does not count against the conceptual point that one can generically derive some field $T^{ab}$ in Newtonian theories just as in GR.

As a final observation, note that nothing in the above argument turns on the derivative operator appearing in the classical spacetime be curved, nor that it arise via the geometrized Poisson equation. In this sense, the argument applies equally well in ordinary Newtonian gravitation as in GNG, just as the argument in section \ref{GR} applies equally well to SR.  But note, too, that the argument applies (only) to situations in which bodies are free in the sense that no further fields appear in their Lagrangian densities, and thus in the ordinary Newtonian case, only when gravitational effects are neglected (since otherwise, a gravitational field would appear in the dynamics for matter).  In other words, in the Newtonian case, in the presence of gravitation, we should expect the mass-momentum of matter alone to have non-vanishing divergence, reflecting the fact that bodies accelerate due to gravity.\footnote{Again, see \citet{Dewar+Weatherall}.}  What is interesting to observe, however, is that if one considered a joint Lagrangian density for matter and the gravitational field, one would expect the argument to go through in such a way that the \emph{total} mass-momentum tensor, for both gravitation and matter, to be divergence-free.

\section{Geometry Revisited}\label{conclusion}

The foregoing discussion might be summarized as follows.  In a wide range of theories, including GR but also GNG, unimodular gravity, TeVeS, and even SR and ordinary Newtonian gravitation, inertial motion, in the form of a geodesic principle, may be understood as a theorem concerning the small-body limit of solutions to the equations governing matter.  These theorems require, as one of their assumptions, that associated with generic matter, there is a symmetric rank-2 tensor $T^{ab}$ that is suitably positive and covariantly divergence-free relative to some derivative operator; the theorems then ensure that sufficiently small bodies (approximately) follow geodesics of this derivative operator.  This means that establishing that small bodies respect the inertial structure encoded by a given derivative operator $\nabla$ requires one to establish that the $T^{ab}$ field associated with matter is divergence-free, or ``conserved'', with respect to $\nabla$.  This condition, in turn, holds generically, in both relativistic and classical spacetimes, of any matter fields whose dynamics, in the absence of interactions with other matter, may be described by a Lagrangian density that is an algebraic function of just appropriate spacetime structures, the matter fields, and their covariant derivatives.  Thus, inertial motion follows under very general conditions from the role that a metric and covariant derivative operator play in the differential equations governing that matter.\footnote{\label{Knox1} The view just expressed is related to, but interestingly distinct from, the ``spacetime functionalism'' defended by \citet{KnoxEffective,KnoxFunctionalism}.  On Knox's view, spacetime is functional in the sense that ``A structure that plays the spacetime role in a theory just is spacetime; once one has analysed the role and understood what fills it, there are no further questions to be asked about the `real' spacetime structure. Spacetime is spacetime by virtue of what it does, not what it is'' \citep[p. 9]{KnoxFunctionalism}; she goes on to clarify that the ``spacetime role is played by whatever defines a structure of local inertial frames'' \citep[p. 10]{KnoxFunctionalism}.  I, meanwhile, argue that inertial motion (which is not to say anything of inertial frames---a notion I find vexed and unhelpful in curved spacetime) follows from the role that the metric and derivative operator play in the dynamics of matter.  In a sense, these views are complementary, then: Knox says that whatever determines inertial structure must represent spacetime, and I show how the metric and covariant derivative operator determine inertial structure.  Still, I would remark on two important differences.  The first is that I am not attempting to give an account of what the ``spacetime concept'' is: on my view, there is no univocal ``spacetime concept'', and in particular, in addition to inertial structure, one might consider causal structure and structure concerning lengths and duration to be essential to any proper understanding of spacetime, at least in some theories.  My goal instead is to investigate the link between a particular aspect of spatiotemporal structure and the dynamics of matter.  The second difference is that, on the view I am defending, the point is not merely that the metric and derivative operator determine inertial structure; rather, it is that they do so precisely because of the role that they play in the matter dynamics.  And, as I will argue below, one can naturally construe that role as one of determining the physically salient notions of length and angle as respected by evolving matter.  In other words, the metric and covariant derivative operator play the functional role Knox considers, but they do so by virtue of playing yet other roles in the theory---roles that are arguably just as ``spatiotemporal'' as the one Knox emphasizes.}

As I have emphasized throughout, the argument just given does not depend on Einstein's equation or anything else special about GR.  Although Einstein's equation may also be used to establish, in GR, that there is a quantity $T^{ab}$ that is covariantly divergence-free, Einstein's equation plays no role in the argument just given, which applies in a much broader range of theories, including TeVeS, where it is the physical metric that determines inertial structure.  And so it seems misleading to emphasize Einstein's equation in discussions of inertial motion in GR.  Ultimately Einstein's equation is not necessary; nor is it sufficient, insofar as it does not establish the energy condition, does not clearly bear on test matter, and cannot establish the relationship between conservation and non-interaction.  Studying the properties of the actual dynamics of matter as a route to conservation, meanwhile, does establish conservation, both for test matter and for non-interacting matter (even when other matter is present).  Although I do not discuss the energy conditions in detail in the present paper, it also seems clear that the most promising route to establishing that an energy conditions holds is, once again, to study the properties of energy-momentum tensors associated with matter, and not via a study of Einstein's equation.

On some level, this conclusion should be congenial to the defender of a ``dynamical perspective'' on spatiotemporal structure such as Brown, who emphasizes the role of the dynamics of matter in underwriting our claims about space and time, since after all, on the view just described inertial motion follows from, and is explained by, the details of the dynamics of matter.  And indeed, it is hard to see how it could be otherwise.\footnote{Here I echo the perspective taken by \citet{Myrvold} (albeit independently).}  Insofar as the coordinating principles that we use to give operational significance to the spacetime metric in GR and other theories---principles like the geodesic principle, the clock hypothesis,\footnote{For a discussion of the clock hypothesis, see \citet{FletcherCH}.  As he shows there, one may view the clock hypothesis as a theorem of GR, at least as regards periodic processes involving null curves (such as so-called ``light clocks'').} and so forth---concern the behavior of matter under idealized circumstances, it had better be the case that the equations actually governing the dynamics of matter are compatible with bodies actually doing what the interpretive principles say they must.  From this perspective it is also fair to say that, as Brown argues in \emph{Physical Relativity}, spacetime structures such as the metric may be viewed as ``a codification of certain key aspects of the behaviour of particles and fields'' (p. 142), at least as regards the link between free, small-body motion and the privileged class of curves picked out by a metric and/or derivative operator.\footnote{See also \citet{Brown+Pooley}.}

And yet, despite this apparent congeniality, there are clear tensions between the conclusions I reach here and Brown's views on inertial motion.  The obvious tension, of course, concerns Brown's claims that GR is starkly different from Newtonian theories and SR with regards to inertial motion.  It is hard to see on what grounds that distinction can be drawn---especially if the difference is meant to come down to the status of the geodesic principle as a theorem.  One is left wondering if what really motivates Brown's claims about inertia is not that the geodesic principle is a theorem in GR, but that the metric and associated derivative operator in GR are dynamical.  This, at least, is what \citet{Sus}, in defending Brown's views, argues makes the explanation of inertia in GR ``more dynamical'' than in other theories.\footnote{Like Brown, Sus also emphasizes the role of Einstein's equation in establishing conservation.}  (A similar suggestion is floated as a reading of Brown by \citet{Samaroo}, who discusses the role of an ``action-reaction'' principle in Brown's arguments.)  But that spacetime structures are dynamical does not distinguish GR from other theories, including unimodular gravity, TeVeS, and even, to a lesser degree, GNG.  And as we have seen, one can see inertial motion as a theorem in these theories---but only if one justifies the conservation condition via the dynamics of matter, in which case the arguments seem to go through in just the same way in theories where spacetime structures are not dynamical.  And so, that the metric is dynamical in GR appears to be irrelevant to the status of inertial motion in the theory.

A related tension is that, from the perspective developed here, it is hard to see anything distinctively ``conspiratorial'' about inertial motion in SR or Newtonian gravitation.  In these theories, just as in GR and TeVeS, inertial motion is a special case of the dynamics of matter, applicable in the small body limit (and more generally, at least in flat spacetime) and in the absence of interactions.  If inertial motion is conspiratorial, then surely \emph{all} motion is conspiratorial.  After all, how are different bodies of a given sort---or fields in different regions---to all know to satisfy the same differential equations?  I do not mean to claim that this question is not interesting or legitimate or even philosophically rich.  But it begins to look less like a question about inertia in particular, and more like an expression of Humean skepticism (or at least puzzlement) concerning the metaphysical status of laws of nature more generally.  And if \emph{this} is what is ultimately at issue, then it is hard to see how GR could be said to provide any help at all.  If our goal is to explain how matter in different regions of space and time can somehow all satisfy the same laws of nature, then merely positing yet more laws of nature cannot be to the point.

Still, there is a deeper issue here.  The situation is not just that the laws governing the matter we observe in the universe happen to have geodesic motion as a special, limiting case.  The arguments presented here show much more than that.  They show that \emph{any} equations governing matter that can be derived from a Lagrangian of a certain sort, with certain dependencies on a metric and its associated derivative operator (or, in the classical case, the classical metrics and derivative operator), will have the same small body, non-interacting behavior.  In other words, the reason that a metric (or metrics) and derivative operator are able to codify the behavior of (generic) matter in the way characterized by the geodesic principle is precisely that that metric and derivative operator are the ones that appear in the dynamics of (all) matter in the relevant ways.  And this, I think, is ultimately what is at the heart of the matter.

As I see it, the most perspicuous explication of what one means, or at least what one should mean, by the claim that spacetime has some geometry, represented by a given metric (or metrics) and derivative operator, is precisely that one can express the dynamics of (all) matter in such a way that all inner products are taken relative to that metric and all derivatives are taken relative to that derivative operator.  This is the physical content of the claim that there are facts about distances, angles, and duration: physical processes occur in such a way that changes in a quantity at a time depend on the state of that quantity and those facts about distances, angles, and duration.  And so, one is left with the conclusion that spacetime structures codify certain facts about the behavior of matter because the dynamics of (all) matter is adapted to those spacetime structures, which is just another way of saying that spacetime ``has'' that geometry.

In fact, Brown makes what I take to be a related point, though he expresses it differently.  In his terms, it is the ``strong equivalence principle'' that allows us to give ``chronometric significance'' to the metric $g_{ab}$.  He initially describes this principle in terms of the existence of certain coordinate systems, but then goes on to say that the strong equivalence principle as he understands it contains two principles that he attributes to \citet{Anderson}.  These principles are (1) ``that measurements on \emph{any} physical system will serve (approximately) to determine the \emph{same} [covariant derivative operator] in a given region of spacetime''; and (2) ``only the [derivative operator] determined by [$g_{ab}$], with its Lorentzian signature, appear in the dynamical laws for these systems'' (p. 170).  This way of characterizing things emphasizes the derivative operator over the metric, when in fact both generally appear in the dynamical laws.  But I think the moral is substantially the same: it is because a particular derivative operator (and metric) appear in a particular way in the equations of motion governing matter that we can ultimately interpret those fields geometrically.

Whether one describes this situation by invoking a principle, or else by saying that spacetime has a given geometry (in the sense described above) strikes me as immaterial.  But Brown places great weight on the further claim that attributing a given geometry to spacetime does no work---or at least, that it plays no explanatory role.  For instance, he rejects the view that he attributes to \citet{Friedman}, \citet{Balashov+Janssen}, and others that it is the fact that, in SR, spacetime is Minkowskian that explains why Maxwell's equations and other field dynamics are Lorentz invariant \citep[pp. 132-143]{Brown}.  Later, he writes: \begin{quote}\singlespacing It is true that in both GR and TeVeS all matter fields couple in the same way to a single `metric' field of Lorentzian signature, which ensures local Lorentz covariance; and by coupling minimally, special relativity in its full glory appears to be valid locally.  But to say that these metric fields are space-time itself, or properties of space-time, is simply to re-express this remarkable double claim, not to account for it. (p. 176)\end{quote}  It is this last claim that seems too strong.  Even if asserting that the metric in GR, or the physical metric in TeVeS, represents spacetime geometry is merely to re-express the claim that it is \emph{this} metric and its associated derivative operator that play the particular roles they do in the dynamics of matter, it nonetheless seems that making this observation---however expressed---carries significant explanatory weight.  It tells us that it is the lengths, angles, and so on described by this particular metric that are the physically salient ones, the ones that enter into the dynamics of matter.  That spacetime has this geometry---i.e., that it is this geometry to which matter dynamics are adapted---is what explains why matter moves as it does.

One can come at this same conclusion from another direction.  In order to express a system of differential equations on a manifold, some background geometrical structure needs to be presupposed.\footnote{This is a moral that I take from \citet{Stein}, who argued in particular that at the heart of the dispute between Leibniz and Newton (or really, Clark) was a disagreement about how much background structure was needed to express adequate laws of motion.  Newton believed he needed the structure of what we now call Newtonian spacetime; Leibniz insisted that any (metaphysically?) adequately theory would make use of strictly less.  But as \citet{Weyl} first showed, Newton's laws do not need the full structure of Newtonian spacetime---instead, they need only Galilean spacetime, i.e., a 4-dimensional affine space, with classical metrics.  One can run a similar argument with Maxwell's equations: the standard of rest provided by an aether in Galilean spacetime effectively provides the structure of Newtonian spacetime; one can understand Lorentz, Einstein, and Minkowski to have each made important steps to the recognition that one can get away with merely Minkowski spacetime, which, like Galilean spacetime, has less structure than Newtonian spacetime (but whose structure is neither less nor more than Galilean spacetime) \citep{BarrettSTS, WeatherallVoid, WeatherallSTG}.  The important point in all of this is that laws of motion, including field equations such as Maxwell's equations, only make sense if one has already postulated some geometrical context.  That said, I do not mean to suggest it is (always) clear precisely what structure must be postulated, as is evidenced by the recent (and not so recent) discussions of the spacetime structure needed for Newton's theory, in light of Corollary VI to the laws of motion \citep{SteinPrehistory,DiSalle,Saunders,KnoxNEP,WeatherallSaunders,DewarMaxwell,WallaceEFG,DewarMaxwell,TehRecovery,WeatherallMaxwell}.}  For instance, if one wants to take derivatives, one needs to work in a space on which a derivative operator is defined.  If one wants to take inner products, or relate vectorial quantities with their duals, one needs a metric.  And so on.  All of the equations governing matter in the history of physics presuppose some structure of this sort.

One can try to identify this structure in multiple ways, of course.  For instance, one can work with coordinate free methods and (attempt to) define, in a coordinate-independent manner, the structures that your equations will presuppose and then write the equations referring only to those structures.  Or one could choose some coordinate system---itself a kind of background structure above and beyond the manifold structure---and express your equations using the coordinate derivative operator and some metric expressed in coordinates.  In this latter case, one can then ask how the same equations might be expressed in different coordinate systems, and discover some transformation rules that will generally pick out ``preferred'' coordinates---i.e., coordinates that are suitably adapted to the geometrical structures to which your equations are adapted.\footnote{The connection between coordinate systems, transformation rules, and geometrical structure is explored in more detail by \citet{Wallace}.}   These may be thought of as ``bottom up'' and ``top down'' approaches to identifying spacetime structure, respectively. Perhaps there are other approaches.  But the end result is the same: one arrives at some system of equations, along with an understanding of the background geometry built into those equations.  If the equations are meant to describe the evolution of matter in space and time, it is natural to describe the geometrical structures to which those equations are adapted ``spacetime geometry''.

Starting in this way makes it look as if spacetime geometry is somehow conceptually prior to dynamics, in just the sense that Brown objects to so strenuously.  And there is a real worry here: having asserted that spacetime has some geometry, one is certainly entitled to ask what the physical interpretation of that geometry is supposed to be.  One can give several answers.  If the geometry in question is Lorentzian, then one can describe how the metric characterizes relations of causal connectibility between events in space and time or the behavior of periodic processes that we might think of as idealized clocks, or how the derivative operator determines the trajectories of small bodies.  But one might be dissatisfied with this sort of answer: after all, you have gone to the trouble to define some dynamics for matter, and whether matter actually behaves the way these interpretive principles claim will surely depend  on the details of those dynamics.  And it does so depend.  But at least in the case of geodesic motion, the interpretive principle is vindicated.  More, as I have shown in the foregoing, it is vindicated precisely because the dynamics in question presuppose just the geometrical structures whose significance we go on to explicate by means of the motion of small bodies.

My purpose in retelling the story in this way is not to defend the opposite view, that somehow it is \emph{only} geometry that plays any explanatory role---or even that geometry is prior to dynamics.  Rather, it is to emphasize the ways in which geometry and dynamics constrain and support one another.\footnote{It is this basic view, that the foundations of a physical theory should be understood as a network of mutually supporting and constraining principles, that I have elsewhere called the ``puzzleball view'' \citep{WeatherallLehmkuhl}.}  To posit a given geometry for spacetime is to limit the resources available for expressing dynamical laws; whether a given geometry is suitable for expressing the actual laws is a highly contingent matter.  Meanwhile, given a system of differential equations, one can extract information about physical geometry by studying things about, for instance, causal cones and propagation speeds, the motion o non-interacting, small bodies, and even the symmetry properties of the equations.  It is this geometrical information that ultimately should be interpreted as physical geometry.  And so, geometry constrains dynamics, but so too do dynamics constrain geometry.  The inferences---and the explanations---go in both directions.

\section*{Acknowledgments}
This paper is partially based upon work supported by the National Science Foundation under Grant No. 1331126 and partially based upon work supported by the John Templeton Foundation grant ``Laws, Methods, and Minds in Cosmology''.  I am grateful to Harvey Brown, Jeremy Butterfield, Erik Curiel, John Earman, Sam Fletcher, Bob Geroch, Marc Holman, Michel Janssen, Eleanor Knox, Dennis Lehmkuhl, David Malament, JB Manchak, Oliver Pooley, Ryan Samaroo, Mike Tamir, Bob Wald, and Michel Janssen for conversations and correspondence (over the years!) related to the arguments in this paper.  I am grateful to Harvey Brown, Erik Curiel, Neil Dewar, David Malament, JB Manchak, Simon Saunders, Chris Smeenk, Chris W\"uthrich, and Jingyi Wu for comments on a previous draft.  Thanks, also, to the Physics Interest Group at the Minnesota Center for Philosophy of Science for joining me in a discussion of \citet{Samaroo} in connection with \citet{Brown} and \citet{Janssen}.

\singlespacing

\end{document}